\documentclass[aps, prl, twocolumn]{revtex4-1}

\usepackage {graphicx}
\usepackage {amsmath}
\usepackage {amsfonts}
\usepackage {amssymb}
\usepackage {mathrsfs}
\usepackage {bm}

\usepackage {epstopdf}
\allowdisplaybreaks[4]

\newcommand {\be}{\begin{eqnarray}}
\newcommand {\ee}{\end{eqnarray}}  
\newcommand {\ba}{\be \begin{aligned}} 
\newcommand {\ea}{\end{aligned} \ee}
\newcommand {\bx}{\mathbf{x}}
\newcommand {\bk}{\mathbf{k}}

\begin{document}

\title{Renormalization group study of interaction-driven quantum anomalous Hall and quantum spin Hall phases in quadratic band crossing systems}



\author{James M. Murray}
\author{Oskar Vafek}
\affiliation{National High Magnetic Field Laboratory and Department of Physics, Florida State University, Tallahassee, FL 32306, USA}


\date{\today}

\begin{abstract}
\noindent 	It is shown that topological insulating phases driven by interactions can be realized without the need for spin-orbit coupling or large intersite interaction in a two-dimensional system of spin-$\frac{1}{2}$ fermions with a single pair of parabolically touching bands. Using a weak-coupling, Wilsonian renormalization group procedure, we show that a quantum anomalous Hall phase is realized for Hubbard interaction, while a quantum spin Hall phase is favored for longer-ranged interactions.
\end{abstract}


\maketitle


Topological aspects of condensed matter systems have become the subject of tremendous interest over the last decade. In two-dimensional (2D) systems, topological insulating phases can be understood as generalizations of the quantum Hall effect. The quantum anomalous Hall (QAH) effect, first studied in Ref.~\cite{haldane88} and recently observed experimentally \cite{chang13}, is an integer quantum Hall phase with gapless edge currents that is realized in the absence of any applied magnetic field. The quantum spin Hall (QSH) effect, predicted in \cite{kane05,bernevig08} and first observed in \cite{koenig07}, is an analogue to the QAH effect, but with opposite chiralities for up and down spins, so that the gapless edge modes are spin (rather than charge) currents \footnote{Strictly speaking, the QSH---sometimes referred to as the ``spin flux'' phase---is not simply identical to two copies of the QAH effect, as the former is classified by a $\mathbb{Z}_2$ (rather than $\mathbb{Z}$) topological invariant. See \textit{e.g.} B.~A.~Bernevig and T.~L.~Hughes, \textit{Topological Insulators and Topological Superconductors} (Princeton University Press, 2013).}. 
Most previous proposals for realizing these effects experimentally \cite{qi06,bernevig08,liu08,liu08b,yu10,xu11,chang13,zhang14} have involved quantum well heterostructures used to engineer systems with topologically nontrivial band structures, typically requiring large spin-orbit coupling to create a ``band inversion.'' While much of the focus has been on noninteracting systems, an important question is whether topological phases can instead be induced by interactions between electrons. Although several models have been proposed to realize such phases \cite{raghu08,sun09,uebelacker11}, often topologically nontrivial phases appear only in regions of parameter space that may be difficult to realize in an experimental system, \textit{e.g.}~when intersite Coulomb repulsion is larger than on-site repulsion or in models of spinless fermions. In this Letter, it is shown that such topologically ordered phases can be realized at weak coupling for {\em any} type of electron-electron interactions in a 2D system of fermions with parabolically touching bands. As we explain below, the use of renormalization group (RG) techniques is crucial for obtaining the correct phases, as the interaction terms leading to these phases may not be present in the bare Hamiltonian and therefore cannot be captured within mean field theory. Rather, the terms leading to topologically nontrivial phases at low energies are generated by fluctuating high-energy modes.

Due to the fact that, unlike linearly dispersing Dirac cones, parabolically touching bands have a finite density of states in 2D, it has been suggested that such systems might host interaction-driven topological phases \cite{sun09,chern12,fradkin13,dora14}. (See also related proposals for bilayer graphene, which in the idealized case features two pairs of parabolically touching bands \cite{nandkishore10,zhang11,lemonik12,zhang12}.) Such parabolically touching bands can arise on the checkerboard and kagome lattices \cite{sun09} at 1/2 and 1/3 filling, respectively, as well as on the Lieb lattice \cite{tsai11} and within certain collinear spin density wave states \cite{chern12}. It has also been proposed that they may arise on 2D surfaces of topological crystalline insulators \cite{fu11}. While RG techniques have been used to investigate the formation of ordered phases in the case of weakly interacting spinless fermions with parabolic band touching \cite{sun09}, up until now the case of spin-$\frac{1}{2}$ has been studied only by mean field theory, which may not be reliable in the case that there are multiple competing instabilities.


In this Letter we use a symmetry-based, Wilsonian renormalization group procedure \cite{vafek10,vafek10b,vafek13} to study the possible weak-coupling phase instabilities of this system. The low-energy effective Hamiltonian used to describe the system is given by $H=H_0 + H_\mathrm{int}$, where the noninteracting part introduced by Sun \textit{et al.}~is given by \cite{sun09}
\ba
\label{eq:0108a}
H_0 &= \sum_{|\bk| < \Lambda} \sum_{\sigma=\uparrow\downarrow} \psi^\dagger_{\bk\sigma} \mathcal{H}_0 (\bk) \psi_{\bk\sigma}, \\
\mathcal{H}_0(\bk) &= t_I k^2 1 + 2t_x k_x k_y \sigma_1 + t_z (k_x^2 - k_y^2)\sigma_3.
\ea
The interacting part, which contains all marginal couplings allowed by symmetry, is given by
\be
\label{eq:H_int}
H_\mathrm{int} = \frac{2\pi}{m} \sum_{i=0}^3 g_i \int d^2 x 
	\left( \sum_{\sigma=\uparrow\downarrow}
	\psi^\dagger_\sigma (\bx) \sigma_i \psi_\sigma (\bx) \right)^2.
\ee
Here $\psi_{\bk\sigma}$ has two components, which in the case of a checkerboard lattice correspond to sublattices $A$ and $B$; $\sigma$ denotes electron spin; and $\sigma_i$ are the standard Pauli matrices, with $\sigma_0 = 1$. Note that there is no term $\sim \sigma_2$ in $\mathcal{H}_0 (\bk)$ since such a term would break time-reversal symmetry. The $d$-wave symmetry of the second and third terms in $\mathcal{H}_0 (\bk)$ give rise to a Berry phase winding of $\pm 2\pi$.
Diagonalizing $\mathcal{H}_0 (\bk)$ gives
\be
\label{eq:bands} 
E_\bk^\pm = \frac{\bk^2}{\sqrt{2}m} \left[ \lambda \pm 
	\sqrt{\cos^2\eta \cos^2 2\theta_\bk + \sin^2\eta \sin^2 2\theta_\bk} \right],
\ee
where $m = 1/\sqrt{2(t_x^2 + t_z^2)}$, $\lambda = t_I / \sqrt{t_x^2 + t_z^2}$, $\cos\eta = t_z / \sqrt{t_x^2 + t_z^2}$, and $\sin\eta = t_x / \sqrt{t_x^2 + t_z^2}$. For $|t_I| < \mathrm{min}(|t_x|,|t_z|)$, \eqref{eq:bands} describes one upward and one downward dispersing band, with the two touching parabolically at $\bk=0$. The Hamiltonian is invariant under the $C_{4v}$ point group, which describes the checkerboard lattice, and time reversal symmetry. The parabolic band touching is stable to perturbations that do not break these symmetries. The group representations and corresponding symmetry-allowed interaction terms $g_i$ are shown in Table \ref{table:matrices}. If the lattice point group is instead $C_{6v}$, as occurs for instance on the kagome lattice, then one must have $g_3 = g_1$ and $|t_x| = |t_z|$ in \eqref{eq:0108a} and \eqref{eq:H_int}. In this case the low-energy effective theory becomes rotationally invariant. The system also exhibits particle-hole symmetry when $\lambda=0$.

\begin{figure}
\centering
\includegraphics[width=0.48\textwidth]{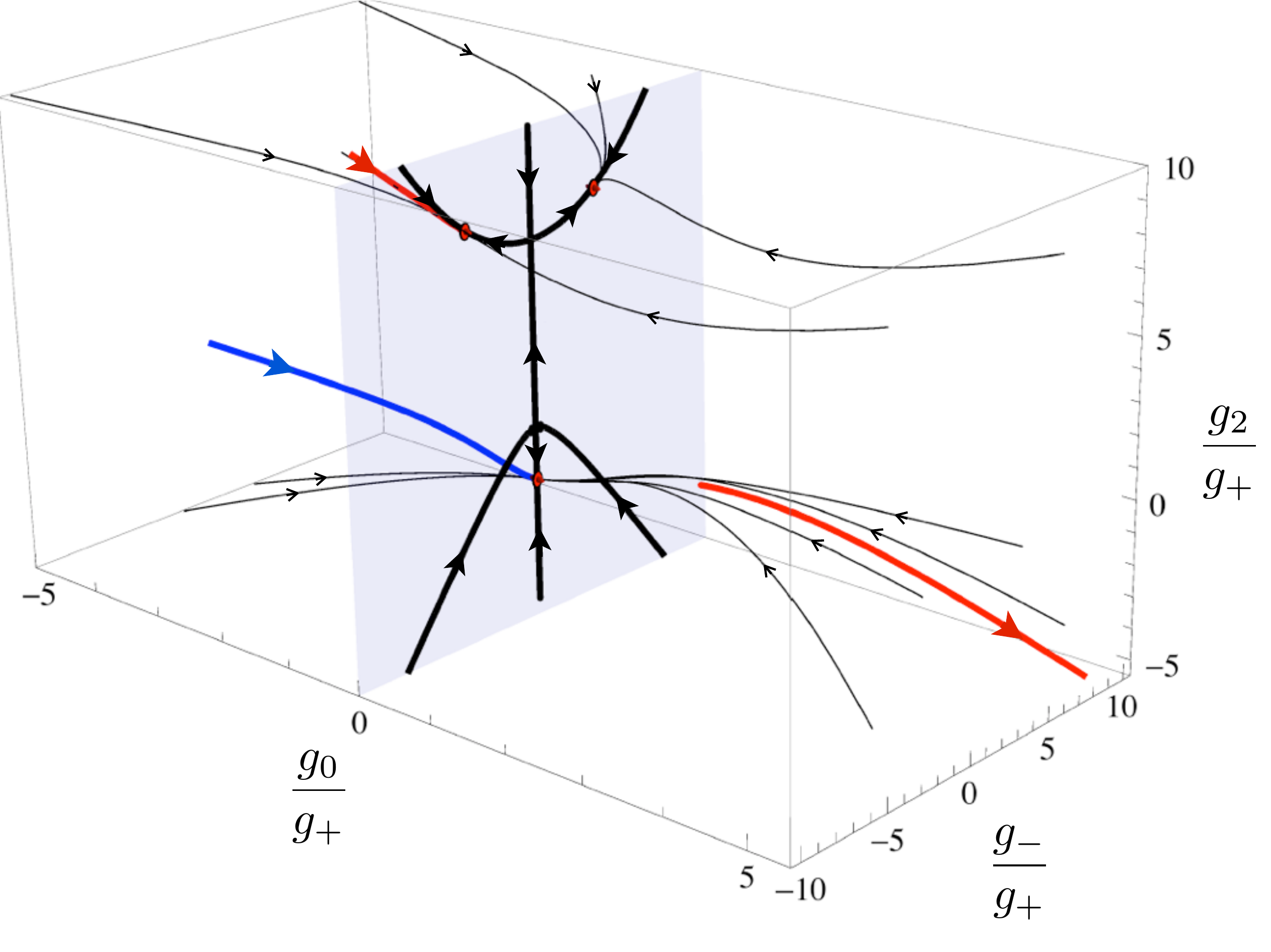}
\caption{Renormalization group flows of the coupling ratios, with $g_\pm = (g_3 \pm g_1)/2$, with fermion dispersion taken to be rotationally invariant ($\eta = \pi/4$) and particle-hole symmetric ($\lambda = 0$). The three stable fixed ratios are shown in red, with the upper two corresponding to a QAH phase, while the stable fixed ratio with $g_2 / g_+ < 0$ corresponds to a QSH phase. The trajectories corresponding to Hubbard and forward scattering interaction are shown as bold red and blue lines, respectively. In the Hubbard case, the flow begins with $g_0 / g_+ > 0$ and then reappears in the opposite quadrant when $g_+(\ell)$ passes through zero. Apart from the unstable Gaussian fixed point, all flows terminate in the $g_0/g_+ = 0$ plane as $g_+ \to -\infty$. For the black arrows, the flow direction corresponds to $g_+ < 0$. If $g_+ > 0$ initially, the couplings first flow opposite to the direction shown (as shown for the Hubbard case) until $g_+$ changes sign and the trajectories follow the arrows shown.
\label{flows3D}}
\end{figure}

We employ a Wilsonian RG procedure in order to study the effects of interactions and instabilities to ordered phases at low energy scales. It is useful to define the following action:
\be
\label{eq:action}
S = \int d\tau \left[ \sum_{|\bk| < \Lambda} \sum_\sigma 
	\psi^\dagger_{\bk\sigma}(\partial_\tau + \mathcal{H}_0 (\bk) )\psi_{\bk\sigma}
	+ H_\mathrm{int} \right],
\ee
where the Grassmann fields $\psi_{\bk\sigma}$ now depend on imaginary time $\tau$.
The RG step is then performed by eliminating states within the momentum shell $\Lambda(1-d\ell) < |\bk| < \Lambda$ while integrating over all frequencies. By including all one-loop diagrams and rescaling the couplings after each RG step, one obtains the following flow equations:
\be
\label{eq:0128a}
\frac{d g_i}{d\ell} = \sum_{j,k=0}^3 A_{ijk} g_j g_k,
\ee
where the coefficients $A_{ijk}$ are given in the Supplementary Information (SI). The parameters $\lambda$ and $\eta$ do not flow at this order. From \eqref{eq:0128a}, one sees that the couplings are marginally relevant and generally flow to infinite values for sufficiently large $\ell$. The {\em ratios} of these couplings, however, approach fixed finite values, with each of these fixed ratios corresponding to a particular ordered phase. Due to the perturbative nature of our approach, the flow equations remain valid only at weak coupling and break down at RG scales where $g_i(\ell)\gtrsim 1$. It is convenient to define the new couplings $g_\pm = (g_3 \pm g_1)/2$. In the limit of a rotationally invariant ($\eta=\pi/4$), particle-hole symmetric ($\lambda=0$) system, the flow equations \eqref{eq:0128a} take on the following relatively simple form (see SI for the case of general $\eta$ and $\lambda$):
\ba
\label{eq:g_flows}
\dot{g}_0 &= -4 g_0 g_+ \\
\dot{g}_+ &= -(g_0 - g_+)^2 - (g_2 - g_+)^2  - 6g_+^2 \\
\dot{g}_2 &= 4(g_0 g_2 - g_2^2 - g_-^2 + g_+^2 - 3 g_2 g_+) \\
\dot{g}_- &= 2 g_-(g_0 - 3g_2 - 2g_+).
\ea
From this equation we see that $g_-$ will not be generated if it vanishes initially, which indeed must be the case when the system has $C_6$ rotational symmetry. One also sees from \eqref{eq:g_flows} that $\dot{g}_+ \leq 0$, indicating that $g_+ (\ell)$ decreases under RG flow. Apart from certain fine-tuned initial conditions from which the coupling ratios flow toward one of the mixed-stability fixed ratios in the $g_0=0$ plane shown in Figure \ref{flows3D}, one finds in all cases that $g_+ (\ell)$ passes through 0 and eventually flows toward large negative values. 
Following the method of Ref.~\cite{vafek10b}, we can take advantage of the monotonic decrease of $g_+(\ell)$ and reparameterize the flow equations in terms of this variable, so that the new flow equations are of the form $d(g_i / g_+)/dg_+ = \Phi_i(\{g_j / g_+ \})$. 

As shown in Figure \ref{flows3D}, which illustrates the RG flows of the coupling ratios parametrized in this way, there are three stable fixed values of the ratios. 
All of the fixed ratios lie in the plane $g_0 / g_+ = 0$. From analyzing susceptibilities (see below), we find that two of the three stable fixed points correspond to the QAH phase, while the third corresponds to QSH. We thus conclude that {\em all} possible instabilities of the system at weak coupling are to topologically ordered phases, and that these instabilities are realized for arbitrarily weak values of the couplings $g_i$. It is useful to describe the flows for various values of the initial couplings. 
For density-density interactions, the only nonzero bare couplings are $g_0$ and $g_3$, with the other couplings generated upon running the RG. For on-site Hubbard interaction on the checkerboard lattice, $g_0 (\ell=0) = g_3 (\ell=0)$, while for the case of long-ranged forward scattering interaction, only $g_0$ is nonzero initially. The spatial range of the interaction can then be adjusted by interpolating between these two limits. For all sufficiently short-ranged interactions satisfying $g_3 (0)/ g_0 (0) > 0.26$, the couplings flow to the fixed ratios at $(g_0, g_-, g_2) = (0, -3.73, 7.46) g_+$, corresponding to the QAH phase. For $g_3 (\ell=0)/ g_0 (\ell=0) < 0.26$, on the other hand, the couplings flow to $(g_0, g_-, g_2) = (0, 0, -1.09) g_+$, corresponding to QSH. Although we have focused on the symmetric case with $\eta = \pi/4$ and $\lambda=0$, the results remain qualitatively similar away from the particle-hole symmetric and rotationally invariant limit.

\begin{figure}
\centering
\includegraphics[width=0.48\textwidth]{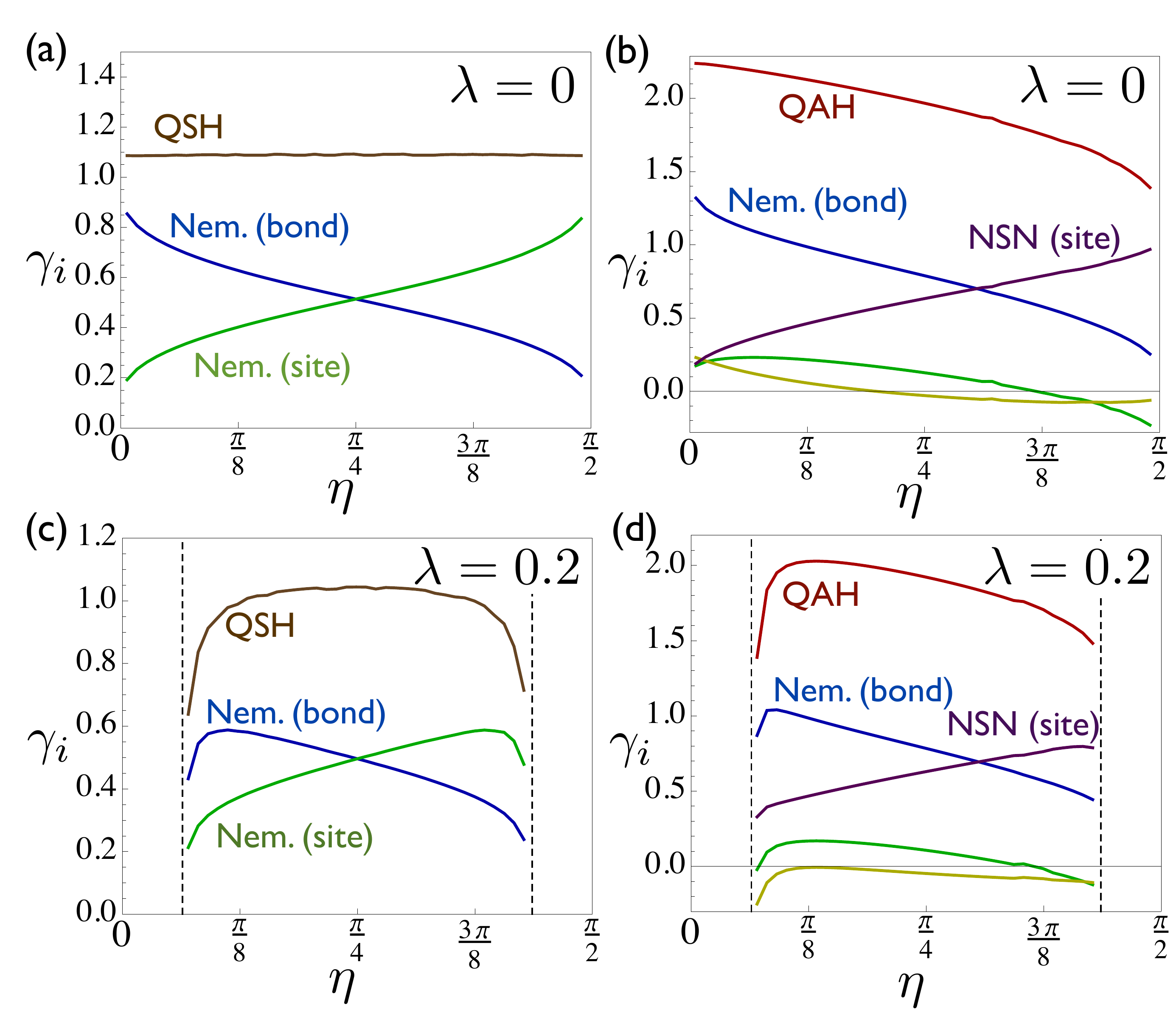}
\caption{Critical exponents of divergent susceptibilities as a function of anisotropy ($\eta=\pi/4$ corresponds to the rotationally invariant case). Left panels correspond to long-ranged interaction ($g_0 (0) >0$ only); right panels correspond to Hubbard interaction ($g_0(0) = g_3(0) > 0$). Upper panels are calculated with particle-hole symmetry; lower panels are calculated with particle-hole asymmetry $\lambda \neq 0$.
\label{gammas}}
\end{figure}

\begin{figure}
\centering
\includegraphics[width=0.45\textwidth]{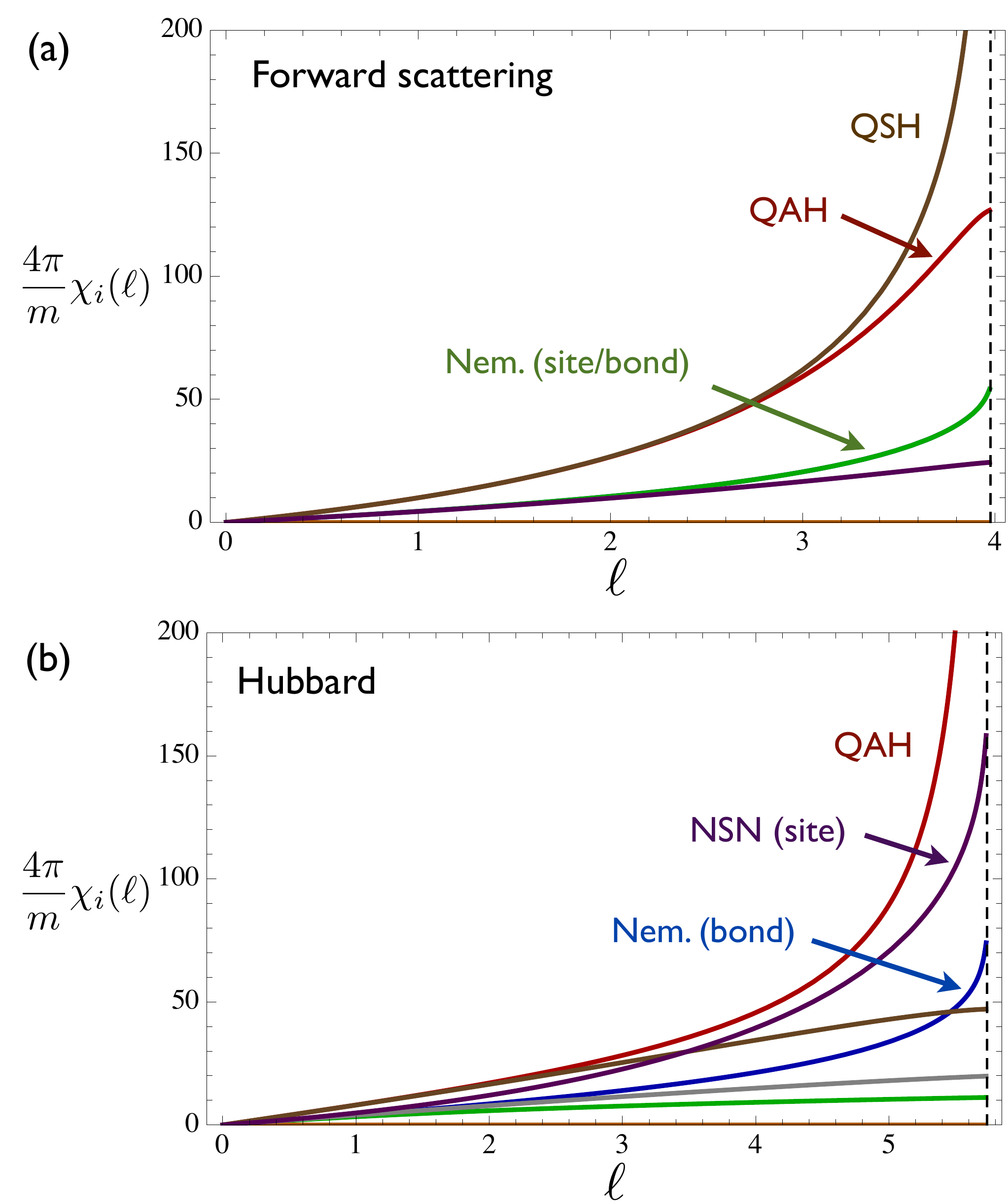}
\caption{Susceptibilities to particle-hole phases as a function of the RG flow parameter $\ell$ calculated for forward scattering (a)  and on-site Hubbard (b) interaction. The electron dispersion is rotationally invariant ($\eta = \pi/4$) and particle-hole symmetric ($\lambda = 0$).
\label{susceptibilities2}}
\end{figure}

In order to investigate possible types of symmetry breaking, we introduce the following source terms into the action:
\ba
\label{eq:0206a}
&S_\Delta = \int d\tau \int d^2 x \bigg\{ \sum_{i=1}^4 \bigg[ \Delta_i^{(c)} \psi^\dagger M_i^{(c)} \psi
	+ \vec{\Delta}_i^{(s)} \cdot \psi^\dagger \mathbf{M}_i^{(s)} \psi \bigg] \\
	&+ \frac{1}{2} \bigg[ \sum_{i=1}^3 \Delta^{(pp)}_i \psi^\dagger M^{(pp)}_i \psi^* 
	+ \vec{\Delta}^{(pp)}_4 \cdot \psi^\dagger \mathbf{M}^{(pp)}_4 \psi^* + H.c. \bigg] \bigg\}.
\ea
The matrices that define the various fermion bilinears in charge (c), spin (s), and particle-particle (pp) channels are given in Table \ref{table:matrices}.
\begin{table}
    \begin{tabular}{lllllll}
    \hline \hline
 Rep. & $g_i$ & $M_i^{(c)}$ & Phase (c) & $M_i^{(s)}$ & Phase (s) & $ M^{(pp)}_i$ \\ \hline
 $A_1$ & $g_0$ & $1_4$ & -- & $1 \vec{s}$ & FM & $1 s_2$ \\
 $A_2$ & $g_2$ & $\sigma_2 1$ & QAH & $\sigma_2 \vec{s}$ & QSH & $\sigma_2 (is_2\vec{s})$ \\
 $B_1$ & $g_3$ & $\sigma_3 1$ & Nem.~(site) & $\sigma_3 \vec{s}$ & NSN (site) & $\sigma_3 s_2$ \\
 $B_2$ & $g_1$ & $\sigma_1 1$ & Nem.~(bond) & $\sigma_1 \vec{s}$ & NSN (bond) & $\sigma_1 s_2$ \\
    \hline \hline
    \end{tabular}
    \caption{Fermionic couplings $g_i$, together with the representation of $C_{4v}$ under which they transform, the matrices appearing in the source term bilinears \eqref{eq:0206a}, and the phases associated with each bilinear. The possible excitonic phases are ferromagnet (FM), quantum anomalous Hall (QAH), quantum spin Hall (QSH), charge nematic on sites or bonds, and nematic-spin-nematic (NSN) on sites or bonds.}
    \label{table:matrices}
\end{table}
The source terms in \eqref{eq:0206a} flow under RG as follows:
\be
\label{eq:0206h}
\frac{d \ln \Delta_i^{(c,s, pp)}}{d\ell} = 2 + \sum_{j=0}^3 B_{ij}^{(c,s,pp)} g_j,
\ee
where the coefficients $B_{ij}^{c,s,pp}$ are provided in the SI. It is then possible to compute susceptibilities by taking derivatives of the free energy with respect to these source terms: $\chi_i = -\partial^2 f /\partial \Delta_i \partial \Delta_i^*$. The full expressions for $\chi_i$ are given in the SI. One finds that they exhibit power law behavior near the RG scale $\ell_*$ where the couplings $g_i(\ell)$ diverge, i.e.~$\chi_i \sim (\ell_* - \ell)^{-\lambda}$. These exponents are given by
\be
\label{eq:0220g}
\gamma_m^{(c,s,pp)} = \frac{2 \sum_j B_{mn}^{(c,s,pp)} \rho_n}{\sum_{ijk} A_{ijk} \rho_i \rho_j \rho_k},
\ee
where $\rho_i = \lim_{\ell\to\ell_*} g_i (\ell) / \sqrt{\sum_j g_j ^2(\ell)}$. The susceptibility exponents are shown for various values of rotational anisotropy and particle-hole asymmetry in Figure \ref{gammas}.
In the case of long-ranged interaction, one can see from the figures that QSH is the leading instability, with subleading instabilities to charge nematic phases. For Hubbard interaction, the leading instability is to the QAH phase, with subleading instabilities to either charge nematic along bonds or nematic spin nematic (NSN) on sites.
The susceptibilities themselves for both types of interaction are shown in Figure \ref{susceptibilities2}, from which we see that the results for the leading instabilities match those from Figure \ref{gammas}. 

The realization of the QAH phase for Hubbard interaction is in apparent contrast with Refs.~\cite{sun09} and \cite{uebelacker11}, both of which found NSN as the leading instability. The NSN phase preserves rotational symmetry in the charge channel while breaking it in the spin channel \cite{oganesyan01,wu07}. The analysis of Ref.~\cite{sun09} was based on a mean-field analysis, which may not be reliable in cases where there are competing instabilities with diverging susceptibilities in multiple channels. Indeed, since the Hubbard interaction does not couple neighboring sites on opposite sublattices, it is clear that the QAH order parameter $\langle \psi^\dagger (\sigma_2 1) \psi \rangle$ will not be favored at mean-field level. The realization of the QAH phase depends crucially on the coupling $g_2$ that is generated by fluctuations that are captured in the RG approach. On the other hand, while the fRG study of Ref.~\cite{uebelacker11} does include such fluctuations, that method is restricted to strong or intermediate couplings $U \gtrsim t$, and so complements our analysis of weak-coupling instabilities. That a NSN phase is realized at strong coupling is not surprising, as this corresponds to a Ne\'el state in which spins anti-align with their nearest neighbors. Such a state is not favored at weak coupling, however, due to its gapless excitation spectrum, which can be expected to gain less condensation energy than the fully gapped QAH phase. As pointed out previously \cite{sun09}, in the charge (spin) nematic phases, the quadratic band touching splits into two (four) Dirac cones, so that the excitation spectrum remains gapless in each of these cases. As the magnitude of the order parameter grows, these cones move further apart. In a theory that takes the full lattice into account, the cones eventually annihilate far from the original band touching points, and the spectrum becomes fully gapped for sufficiently strong interactions. At weak coupling, however, this cannot occur, and one expects the fully gapped spectra of the QAH or QSH phases to be favored in this case, which is indeed what we find. The study of the quantum phase transition between the QAH and NSN phases is an interesting possible direction for future work.

In conclusion, we have shown using a weak-coupling RG framework that it is possible to realize topological QAH and QSH phases from electron-electron interactions in a system of parabolically touching bands. The fact that these phases are realized for arbitrarily weak interactions of any type suggests that they are likely to be observed experimentally if 2D materials with a single quadratic band crossing point can be realized.

This work was supported by the NSF CAREER award
under Grant No. DMR-0955561 (OV), NSF Cooperative
Agreement No. DMR-0654118, and the State of Florida
(OV,JM).

\bibliography{biblio}

\begin{thebibliography}{29}%
\makeatletter
\providecommand \@ifxundefined [1]{%
 \@ifx{#1\undefined}
}%
\providecommand \@ifnum [1]{%
 \ifnum #1\expandafter \@firstoftwo
 \else \expandafter \@secondoftwo
 \fi
}%
\providecommand \@ifx [1]{%
 \ifx #1\expandafter \@firstoftwo
 \else \expandafter \@secondoftwo
 \fi
}%
\providecommand \natexlab [1]{#1}%
\providecommand \enquote  [1]{``#1''}%
\providecommand \bibnamefont  [1]{#1}%
\providecommand \bibfnamefont [1]{#1}%
\providecommand \citenamefont [1]{#1}%
\providecommand \href@noop [0]{\@secondoftwo}%
\providecommand \href [0]{\begingroup \@sanitize@url \@href}%
\providecommand \@href[1]{\@@startlink{#1}\@@href}%
\providecommand \@@href[1]{\endgroup#1\@@endlink}%
\providecommand \@sanitize@url [0]{\catcode `\\12\catcode `\$12\catcode
  `\&12\catcode `\#12\catcode `\^12\catcode `\_12\catcode `\%12\relax}%
\providecommand \@@startlink[1]{}%
\providecommand \@@endlink[0]{}%
\providecommand \url  [0]{\begingroup\@sanitize@url \@url }%
\providecommand \@url [1]{\endgroup\@href {#1}{\urlprefix }}%
\providecommand \urlprefix  [0]{URL }%
\providecommand \Eprint [0]{\href }%
\providecommand \doibase [0]{http://dx.doi.org/}%
\providecommand \selectlanguage [0]{\@gobble}%
\providecommand \bibinfo  [0]{\@secondoftwo}%
\providecommand \bibfield  [0]{\@secondoftwo}%
\providecommand \translation [1]{[#1]}%
\providecommand \BibitemOpen [0]{}%
\providecommand \bibitemStop [0]{}%
\providecommand \bibitemNoStop [0]{.\EOS\space}%
\providecommand \EOS [0]{\spacefactor3000\relax}%
\providecommand \BibitemShut  [1]{\csname bibitem#1\endcsname}%
\let\auto@bib@innerbib\@empty
\bibitem [{\citenamefont {Haldane}(1988)}]{haldane88}%
  \BibitemOpen
  \bibfield  {author} {\bibinfo {author} {\bibfnamefont {F.~D.~M.}\
  \bibnamefont {Haldane}},\ }\href@noop {} {\bibfield  {journal} {\bibinfo
  {journal} {Physical Review Letters}\ }\textbf {\bibinfo {volume} {61}},\
  \bibinfo {pages} {2015} (\bibinfo {year} {1988})}\BibitemShut {NoStop}%
\bibitem [{\citenamefont {Chang}\ \emph {et~al.}(2013)\citenamefont {Chang},
  \citenamefont {Zhang}, \citenamefont {Feng}, \citenamefont {Shen},
  \citenamefont {Zhang}, \citenamefont {Guo}, \citenamefont {Li}, \citenamefont
  {Ou}, \citenamefont {Wei}, \citenamefont {Wang}, \citenamefont {Ji},
  \citenamefont {Feng}, \citenamefont {Ji}, \citenamefont {Chen}, \citenamefont
  {Jia}, \citenamefont {Dai}, \citenamefont {Fang}, \citenamefont {Zhang},
  \citenamefont {He}, \citenamefont {Wang}, \citenamefont {Lu}, \citenamefont
  {Ma},\ and\ \citenamefont {Xue}}]{chang13}%
  \BibitemOpen
  \bibfield  {author} {\bibinfo {author} {\bibfnamefont {C.-Z.}\ \bibnamefont
  {Chang}}, \bibinfo {author} {\bibfnamefont {J.}~\bibnamefont {Zhang}},
  \bibinfo {author} {\bibfnamefont {X.}~\bibnamefont {Feng}}, \bibinfo {author}
  {\bibfnamefont {J.}~\bibnamefont {Shen}}, \bibinfo {author} {\bibfnamefont
  {Z.}~\bibnamefont {Zhang}}, \bibinfo {author} {\bibfnamefont
  {M.}~\bibnamefont {Guo}}, \bibinfo {author} {\bibfnamefont {K.}~\bibnamefont
  {Li}}, \bibinfo {author} {\bibfnamefont {Y.}~\bibnamefont {Ou}}, \bibinfo
  {author} {\bibfnamefont {P.}~\bibnamefont {Wei}}, \bibinfo {author}
  {\bibfnamefont {L.-L.}\ \bibnamefont {Wang}}, \bibinfo {author}
  {\bibfnamefont {Z.-Q.}\ \bibnamefont {Ji}}, \bibinfo {author} {\bibfnamefont
  {Y.}~\bibnamefont {Feng}}, \bibinfo {author} {\bibfnamefont {S.}~\bibnamefont
  {Ji}}, \bibinfo {author} {\bibfnamefont {X.}~\bibnamefont {Chen}}, \bibinfo
  {author} {\bibfnamefont {J.}~\bibnamefont {Jia}}, \bibinfo {author}
  {\bibfnamefont {X.}~\bibnamefont {Dai}}, \bibinfo {author} {\bibfnamefont
  {Z.}~\bibnamefont {Fang}}, \bibinfo {author} {\bibfnamefont {S.-C.}\
  \bibnamefont {Zhang}}, \bibinfo {author} {\bibfnamefont {K.}~\bibnamefont
  {He}}, \bibinfo {author} {\bibfnamefont {Y.}~\bibnamefont {Wang}}, \bibinfo
  {author} {\bibfnamefont {L.}~\bibnamefont {Lu}}, \bibinfo {author}
  {\bibfnamefont {X.-C.}\ \bibnamefont {Ma}}, \ and\ \bibinfo {author}
  {\bibfnamefont {Q.-K.}\ \bibnamefont {Xue}},\ }\href {\doibase
  10.1126/science.1234414} {\bibfield  {journal} {\bibinfo  {journal}
  {Science}\ }\textbf {\bibinfo {volume} {340}},\ \bibinfo {pages} {167}
  (\bibinfo {year} {2013})}\BibitemShut {NoStop}%
\bibitem [{\citenamefont {Kane}\ and\ \citenamefont {Mele}(2005)}]{kane05}%
  \BibitemOpen
  \bibfield  {author} {\bibinfo {author} {\bibfnamefont {C.~L.}\ \bibnamefont
  {Kane}}\ and\ \bibinfo {author} {\bibfnamefont {E.~J.}\ \bibnamefont
  {Mele}},\ }\href {\doibase 10.1103/PhysRevLett.95.146802} {\bibfield
  {journal} {\bibinfo  {journal} {Phys. Rev. Lett.}\ }\textbf {\bibinfo
  {volume} {95}},\ \bibinfo {pages} {146802} (\bibinfo {year}
  {2005})}\BibitemShut {NoStop}%
\bibitem [{\citenamefont {Bernevig}\ \emph {et~al.}(2006)\citenamefont
  {Bernevig}, \citenamefont {Hughes},\ and\ \citenamefont
  {Zhang}}]{bernevig08}%
  \BibitemOpen
  \bibfield  {author} {\bibinfo {author} {\bibfnamefont {B.~A.}\ \bibnamefont
  {Bernevig}}, \bibinfo {author} {\bibfnamefont {T.~L.}\ \bibnamefont
  {Hughes}}, \ and\ \bibinfo {author} {\bibfnamefont {S.-C.}\ \bibnamefont
  {Zhang}},\ }\href@noop {} {\bibfield  {journal} {\bibinfo  {journal}
  {Science}\ }\textbf {\bibinfo {volume} {314}},\ \bibinfo {pages} {1757}
  (\bibinfo {year} {2006})}\BibitemShut {NoStop}%
\bibitem [{\citenamefont {K\"onig}\ \emph {et~al.}(2007)\citenamefont
  {K\"onig}, \citenamefont {Wiedmann}, \citenamefont {Br\"une}, \citenamefont
  {Roth}, \citenamefont {Buhmann}, \citenamefont {Molenkamp}, \citenamefont
  {Qi},\ and\ \citenamefont {Zhang}}]{koenig07}%
  \BibitemOpen
  \bibfield  {author} {\bibinfo {author} {\bibfnamefont {M.}~\bibnamefont
  {K\"onig}}, \bibinfo {author} {\bibfnamefont {S.}~\bibnamefont {Wiedmann}},
  \bibinfo {author} {\bibfnamefont {C.}~\bibnamefont {Br\"une}}, \bibinfo
  {author} {\bibfnamefont {A.}~\bibnamefont {Roth}}, \bibinfo {author}
  {\bibfnamefont {H.}~\bibnamefont {Buhmann}}, \bibinfo {author} {\bibfnamefont
  {L.~W.}\ \bibnamefont {Molenkamp}}, \bibinfo {author} {\bibfnamefont {X.-L.}\
  \bibnamefont {Qi}}, \ and\ \bibinfo {author} {\bibfnamefont {S.-C.}\
  \bibnamefont {Zhang}},\ }\href {\doibase 10.1126/science.1148047} {\bibfield
  {journal} {\bibinfo  {journal} {Science}\ }\textbf {\bibinfo {volume}
  {318}},\ \bibinfo {pages} {766} (\bibinfo {year} {2007})}\BibitemShut
  {NoStop}%
\bibitem [{Note1()}]{Note1}%
  \BibitemOpen
  \bibinfo {note} {Strictly speaking, the QSH---sometimes referred to as the
  ``spin flux'' phase---is not simply identical to two copies of the QAH
  effect, as the former is classified by a $\protect \mathbb {Z}_2$ (rather
  than $\protect \mathbb {Z}$) topological invariant. See \protect \textit
  {e.g.} B.~A.~Bernevig and T.~L.~Hughes, \protect \textit {Topological
  Insulators and Topological Superconductors} (Princeton University Press,
  2013).}\BibitemShut {Stop}%
\bibitem [{\citenamefont {Qi}\ \emph {et~al.}(2006)\citenamefont {Qi},
  \citenamefont {Wu},\ and\ \citenamefont {Zhang}}]{qi06}%
  \BibitemOpen
  \bibfield  {author} {\bibinfo {author} {\bibfnamefont {X.-L.}\ \bibnamefont
  {Qi}}, \bibinfo {author} {\bibfnamefont {Y.-S.}\ \bibnamefont {Wu}}, \ and\
  \bibinfo {author} {\bibfnamefont {S.-C.}\ \bibnamefont {Zhang}},\ }\href
  {\doibase 10.1103/PhysRevB.74.085308} {\bibfield  {journal} {\bibinfo
  {journal} {Phys. Rev. B}\ }\textbf {\bibinfo {volume} {74}},\ \bibinfo
  {pages} {085308} (\bibinfo {year} {2006})}\BibitemShut {NoStop}%
\bibitem [{\citenamefont {Liu}\ \emph {et~al.}(2008{\natexlab{a}})\citenamefont
  {Liu}, \citenamefont {Hughes}, \citenamefont {Qi}, \citenamefont {Wang},\
  and\ \citenamefont {Zhang}}]{liu08}%
  \BibitemOpen
  \bibfield  {author} {\bibinfo {author} {\bibfnamefont {C.}~\bibnamefont
  {Liu}}, \bibinfo {author} {\bibfnamefont {T.~L.}\ \bibnamefont {Hughes}},
  \bibinfo {author} {\bibfnamefont {X.-L.}\ \bibnamefont {Qi}}, \bibinfo
  {author} {\bibfnamefont {K.}~\bibnamefont {Wang}}, \ and\ \bibinfo {author}
  {\bibfnamefont {S.-C.}\ \bibnamefont {Zhang}},\ }\href@noop {} {\bibfield
  {journal} {\bibinfo  {journal} {Physical review letters}\ }\textbf {\bibinfo
  {volume} {100}},\ \bibinfo {pages} {236601} (\bibinfo {year}
  {2008}{\natexlab{a}})}\BibitemShut {NoStop}%
\bibitem [{\citenamefont {Liu}\ \emph {et~al.}(2008{\natexlab{b}})\citenamefont
  {Liu}, \citenamefont {Qi}, \citenamefont {Dai}, \citenamefont {Fang},\ and\
  \citenamefont {Zhang}}]{liu08b}%
  \BibitemOpen
  \bibfield  {author} {\bibinfo {author} {\bibfnamefont {C.-X.}\ \bibnamefont
  {Liu}}, \bibinfo {author} {\bibfnamefont {X.-L.}\ \bibnamefont {Qi}},
  \bibinfo {author} {\bibfnamefont {X.}~\bibnamefont {Dai}}, \bibinfo {author}
  {\bibfnamefont {Z.}~\bibnamefont {Fang}}, \ and\ \bibinfo {author}
  {\bibfnamefont {S.-C.}\ \bibnamefont {Zhang}},\ }\href {\doibase
  10.1103/PhysRevLett.101.146802} {\bibfield  {journal} {\bibinfo  {journal}
  {Phys. Rev. Lett.}\ }\textbf {\bibinfo {volume} {101}},\ \bibinfo {pages}
  {146802} (\bibinfo {year} {2008}{\natexlab{b}})}\BibitemShut {NoStop}%
\bibitem [{\citenamefont {Yu}\ \emph {et~al.}(2010)\citenamefont {Yu},
  \citenamefont {Zhang}, \citenamefont {Zhang}, \citenamefont {Zhang},
  \citenamefont {Dai},\ and\ \citenamefont {Fang}}]{yu10}%
  \BibitemOpen
  \bibfield  {author} {\bibinfo {author} {\bibfnamefont {R.}~\bibnamefont
  {Yu}}, \bibinfo {author} {\bibfnamefont {W.}~\bibnamefont {Zhang}}, \bibinfo
  {author} {\bibfnamefont {H.-J.}\ \bibnamefont {Zhang}}, \bibinfo {author}
  {\bibfnamefont {S.-C.}\ \bibnamefont {Zhang}}, \bibinfo {author}
  {\bibfnamefont {X.}~\bibnamefont {Dai}}, \ and\ \bibinfo {author}
  {\bibfnamefont {Z.}~\bibnamefont {Fang}},\ }\href {\doibase
  10.1126/science.1187485} {\bibfield  {journal} {\bibinfo  {journal}
  {Science}\ }\textbf {\bibinfo {volume} {329}},\ \bibinfo {pages} {61}
  (\bibinfo {year} {2010})}\BibitemShut {NoStop}%
\bibitem [{\citenamefont {Xu}\ \emph {et~al.}(2011)\citenamefont {Xu},
  \citenamefont {Weng}, \citenamefont {Wang}, \citenamefont {Dai},\ and\
  \citenamefont {Fang}}]{xu11}%
  \BibitemOpen
  \bibfield  {author} {\bibinfo {author} {\bibfnamefont {G.}~\bibnamefont
  {Xu}}, \bibinfo {author} {\bibfnamefont {H.}~\bibnamefont {Weng}}, \bibinfo
  {author} {\bibfnamefont {Z.}~\bibnamefont {Wang}}, \bibinfo {author}
  {\bibfnamefont {X.}~\bibnamefont {Dai}}, \ and\ \bibinfo {author}
  {\bibfnamefont {Z.}~\bibnamefont {Fang}},\ }\href {\doibase
  10.1103/PhysRevLett.107.186806} {\bibfield  {journal} {\bibinfo  {journal}
  {Phys. Rev. Lett.}\ }\textbf {\bibinfo {volume} {107}},\ \bibinfo {pages}
  {186806} (\bibinfo {year} {2011})}\BibitemShut {NoStop}%
\bibitem [{\citenamefont {Zhang}\ \emph {et~al.}(2014)\citenamefont {Zhang},
  \citenamefont {Xu}, \citenamefont {Wang},\ and\ \citenamefont
  {Zhang}}]{zhang14}%
  \BibitemOpen
  \bibfield  {author} {\bibinfo {author} {\bibfnamefont {H.}~\bibnamefont
  {Zhang}}, \bibinfo {author} {\bibfnamefont {Y.}~\bibnamefont {Xu}}, \bibinfo
  {author} {\bibfnamefont {J.}~\bibnamefont {Wang}}, \ and\ \bibinfo {author}
  {\bibfnamefont {S.-C.}\ \bibnamefont {Zhang}},\ }\href
  {http://arxiv.org/abs/1402.5167} {\  (\bibinfo {year} {2014})},\ \Eprint
  {http://arxiv.org/abs/arXiv.org:1402.5167} {arXiv.org:1402.5167} \BibitemShut
  {NoStop}%
\bibitem [{\citenamefont {Raghu}\ \emph {et~al.}(2008)\citenamefont {Raghu},
  \citenamefont {Qi}, \citenamefont {Honerkamp},\ and\ \citenamefont
  {Zhang}}]{raghu08}%
  \BibitemOpen
  \bibfield  {author} {\bibinfo {author} {\bibfnamefont {S.}~\bibnamefont
  {Raghu}}, \bibinfo {author} {\bibfnamefont {X.-L.}\ \bibnamefont {Qi}},
  \bibinfo {author} {\bibfnamefont {C.}~\bibnamefont {Honerkamp}}, \ and\
  \bibinfo {author} {\bibfnamefont {S.-C.}\ \bibnamefont {Zhang}},\ }\href
  {\doibase 10.1103/PhysRevLett.100.156401} {\bibfield  {journal} {\bibinfo
  {journal} {Phys. Rev. Lett.}\ }\textbf {\bibinfo {volume} {100}},\ \bibinfo
  {pages} {156401} (\bibinfo {year} {2008})}\BibitemShut {NoStop}%
\bibitem [{\citenamefont {Sun}\ \emph {et~al.}(2009)\citenamefont {Sun},
  \citenamefont {Yao}, \citenamefont {Fradkin},\ and\ \citenamefont
  {Kivelson}}]{sun09}%
  \BibitemOpen
  \bibfield  {author} {\bibinfo {author} {\bibfnamefont {K.}~\bibnamefont
  {Sun}}, \bibinfo {author} {\bibfnamefont {H.}~\bibnamefont {Yao}}, \bibinfo
  {author} {\bibfnamefont {E.}~\bibnamefont {Fradkin}}, \ and\ \bibinfo
  {author} {\bibfnamefont {S.~A.}\ \bibnamefont {Kivelson}},\ }\href {\doibase
  10.1103/PhysRevLett.103.046811} {\bibfield  {journal} {\bibinfo  {journal}
  {Phys. Rev. Lett.}\ }\textbf {\bibinfo {volume} {103}},\ \bibinfo {pages}
  {046811} (\bibinfo {year} {2009})}\BibitemShut {NoStop}%
\bibitem [{\citenamefont {Uebelacker}\ and\ \citenamefont
  {Honerkamp}(2011)}]{uebelacker11}%
  \BibitemOpen
  \bibfield  {author} {\bibinfo {author} {\bibfnamefont {S.}~\bibnamefont
  {Uebelacker}}\ and\ \bibinfo {author} {\bibfnamefont {C.}~\bibnamefont
  {Honerkamp}},\ }\href@noop {} {\bibfield  {journal} {\bibinfo  {journal}
  {Physical Review B}\ }\textbf {\bibinfo {volume} {84}},\ \bibinfo {pages}
  {205122} (\bibinfo {year} {2011})}\BibitemShut {NoStop}%
\bibitem [{\citenamefont {Chern}\ and\ \citenamefont
  {Batista}(2012)}]{chern12}%
  \BibitemOpen
  \bibfield  {author} {\bibinfo {author} {\bibfnamefont {G.-W.}\ \bibnamefont
  {Chern}}\ and\ \bibinfo {author} {\bibfnamefont {C.~D.}\ \bibnamefont
  {Batista}},\ }\href {\doibase 10.1103/PhysRevLett.109.156801} {\bibfield
  {journal} {\bibinfo  {journal} {Phys. Rev. Lett.}\ }\textbf {\bibinfo
  {volume} {109}},\ \bibinfo {pages} {156801} (\bibinfo {year}
  {2012})}\BibitemShut {NoStop}%
\bibitem [{\citenamefont {Fradkin}(2013)}]{fradkin13}%
  \BibitemOpen
  \bibfield  {author} {\bibinfo {author} {\bibfnamefont {E.}~\bibnamefont
  {Fradkin}},\ }\href@noop {} {\emph {\bibinfo {title} {Field Theories of
  Condensed Matter Physics}}}\ (\bibinfo  {publisher} {Cambridge University
  Press},\ \bibinfo {year} {2013})\BibitemShut {NoStop}%
\bibitem [{\citenamefont {Dor\'a}\ \emph {et~al.}(2014)\citenamefont {Dor\'a},
  \citenamefont {Herbut},\ and\ \citenamefont {Moessner}}]{dora14}%
  \BibitemOpen
  \bibfield  {author} {\bibinfo {author} {\bibfnamefont {B.}~\bibnamefont
  {Dor\'a}}, \bibinfo {author} {\bibfnamefont {I.~F.}\ \bibnamefont {Herbut}},
  \ and\ \bibinfo {author} {\bibfnamefont {R.}~\bibnamefont {Moessner}},\
  }\href {http://arxiv.org/abs/1402.6532} {\  (\bibinfo {year} {2014})},\
  \Eprint {http://arxiv.org/abs/arXiv.org:1402.6532} {arXiv.org:1402.6532}
  \BibitemShut {NoStop}%
\bibitem [{\citenamefont {Nandkishore}\ and\ \citenamefont
  {Levitov}(2010)}]{nandkishore10}%
  \BibitemOpen
  \bibfield  {author} {\bibinfo {author} {\bibfnamefont {R.}~\bibnamefont
  {Nandkishore}}\ and\ \bibinfo {author} {\bibfnamefont {L.}~\bibnamefont
  {Levitov}},\ }\href {\doibase 10.1103/PhysRevB.82.115124} {\bibfield
  {journal} {\bibinfo  {journal} {Phys. Rev. B}\ }\textbf {\bibinfo {volume}
  {82}},\ \bibinfo {pages} {115124} (\bibinfo {year} {2010})}\BibitemShut
  {NoStop}%
\bibitem [{\citenamefont {Zhang}\ \emph {et~al.}(2011)\citenamefont {Zhang},
  \citenamefont {Jung}, \citenamefont {Fiete}, \citenamefont {Niu},\ and\
  \citenamefont {MacDonald}}]{zhang11}%
  \BibitemOpen
  \bibfield  {author} {\bibinfo {author} {\bibfnamefont {F.}~\bibnamefont
  {Zhang}}, \bibinfo {author} {\bibfnamefont {J.}~\bibnamefont {Jung}},
  \bibinfo {author} {\bibfnamefont {G.~A.}\ \bibnamefont {Fiete}}, \bibinfo
  {author} {\bibfnamefont {Q.}~\bibnamefont {Niu}}, \ and\ \bibinfo {author}
  {\bibfnamefont {A.~H.}\ \bibnamefont {MacDonald}},\ }\href {\doibase
  10.1103/PhysRevLett.106.156801} {\bibfield  {journal} {\bibinfo  {journal}
  {Phys. Rev. Lett.}\ }\textbf {\bibinfo {volume} {106}},\ \bibinfo {pages}
  {156801} (\bibinfo {year} {2011})}\BibitemShut {NoStop}%
\bibitem [{\citenamefont {Lemonik}\ \emph {et~al.}(2012)\citenamefont
  {Lemonik}, \citenamefont {Aleiner},\ and\ \citenamefont
  {Fal'ko}}]{lemonik12}%
  \BibitemOpen
  \bibfield  {author} {\bibinfo {author} {\bibfnamefont {Y.}~\bibnamefont
  {Lemonik}}, \bibinfo {author} {\bibfnamefont {I.}~\bibnamefont {Aleiner}}, \
  and\ \bibinfo {author} {\bibfnamefont {V.~I.}\ \bibnamefont {Fal'ko}},\
  }\href {\doibase 10.1103/PhysRevB.85.245451} {\bibfield  {journal} {\bibinfo
  {journal} {Phys. Rev. B}\ }\textbf {\bibinfo {volume} {85}},\ \bibinfo
  {pages} {245451} (\bibinfo {year} {2012})}\BibitemShut {NoStop}%
\bibitem [{\citenamefont {Zhang}\ \emph {et~al.}(2012)\citenamefont {Zhang},
  \citenamefont {Min},\ and\ \citenamefont {MacDonald}}]{zhang12}%
  \BibitemOpen
  \bibfield  {author} {\bibinfo {author} {\bibfnamefont {F.}~\bibnamefont
  {Zhang}}, \bibinfo {author} {\bibfnamefont {H.}~\bibnamefont {Min}}, \ and\
  \bibinfo {author} {\bibfnamefont {A.}~\bibnamefont {MacDonald}},\ }\href@noop
  {} {\bibfield  {journal} {\bibinfo  {journal} {Physical Review B}\ }\textbf
  {\bibinfo {volume} {86}},\ \bibinfo {pages} {155128} (\bibinfo {year}
  {2012})}\BibitemShut {NoStop}%
\bibitem [{\citenamefont {Tsai}\ \emph {et~al.}(2011)\citenamefont {Tsai},
  \citenamefont {Fang}, \citenamefont {Yao},\ and\ \citenamefont
  {Hu}}]{tsai11}%
  \BibitemOpen
  \bibfield  {author} {\bibinfo {author} {\bibfnamefont {W.-F.}\ \bibnamefont
  {Tsai}}, \bibinfo {author} {\bibfnamefont {C.}~\bibnamefont {Fang}}, \bibinfo
  {author} {\bibfnamefont {H.}~\bibnamefont {Yao}}, \ and\ \bibinfo {author}
  {\bibfnamefont {J.}~\bibnamefont {Hu}},\ }\href
  {http://arxiv.org/abs/1112.5789} {\  (\bibinfo {year} {2011})},\ \Eprint
  {http://arxiv.org/abs/arXiv.org:1112.5789} {arXiv.org:1112.5789} \BibitemShut
  {NoStop}%
\bibitem [{\citenamefont {Fu}(2011)}]{fu11}%
  \BibitemOpen
  \bibfield  {author} {\bibinfo {author} {\bibfnamefont {L.}~\bibnamefont
  {Fu}},\ }\href {\doibase 10.1103/PhysRevLett.106.106802} {\bibfield
  {journal} {\bibinfo  {journal} {Phys. Rev. Lett.}\ }\textbf {\bibinfo
  {volume} {106}},\ \bibinfo {pages} {106802} (\bibinfo {year}
  {2011})}\BibitemShut {NoStop}%
\bibitem [{\citenamefont {Vafek}(2010)}]{vafek10}%
  \BibitemOpen
  \bibfield  {author} {\bibinfo {author} {\bibfnamefont {O.}~\bibnamefont
  {Vafek}},\ }\href {\doibase 10.1103/PhysRevB.82.205106} {\bibfield  {journal}
  {\bibinfo  {journal} {Phys. Rev. B}\ }\textbf {\bibinfo {volume} {82}},\
  \bibinfo {pages} {205106} (\bibinfo {year} {2010})}\BibitemShut {NoStop}%
\bibitem [{\citenamefont {Vafek}\ and\ \citenamefont {Yang}(2010)}]{vafek10b}%
  \BibitemOpen
  \bibfield  {author} {\bibinfo {author} {\bibfnamefont {O.}~\bibnamefont
  {Vafek}}\ and\ \bibinfo {author} {\bibfnamefont {K.}~\bibnamefont {Yang}},\
  }\href {\doibase 10.1103/PhysRevB.81.041401} {\bibfield  {journal} {\bibinfo
  {journal} {Phys. Rev. B}\ }\textbf {\bibinfo {volume} {81}},\ \bibinfo
  {pages} {041401} (\bibinfo {year} {2010})}\BibitemShut {NoStop}%
\bibitem [{\citenamefont {Vafek}\ \emph {et~al.}(2014)\citenamefont {Vafek},
  \citenamefont {Murray},\ and\ \citenamefont {Cvetkovic}}]{vafek13}%
  \BibitemOpen
  \bibfield  {author} {\bibinfo {author} {\bibfnamefont {O.}~\bibnamefont
  {Vafek}}, \bibinfo {author} {\bibfnamefont {J.~M.}\ \bibnamefont {Murray}}, \
  and\ \bibinfo {author} {\bibfnamefont {V.}~\bibnamefont {Cvetkovic}},\ }\href
  {\doibase 10.1103/PhysRevLett.112.147002} {\bibfield  {journal} {\bibinfo
  {journal} {Phys. Rev. Lett.}\ }\textbf {\bibinfo {volume} {112}},\ \bibinfo
  {pages} {147002} (\bibinfo {year} {2014})}\BibitemShut {NoStop}%
\bibitem [{\citenamefont {Oganesyan}\ \emph {et~al.}(2001)\citenamefont
  {Oganesyan}, \citenamefont {Kivelson},\ and\ \citenamefont
  {Fradkin}}]{oganesyan01}%
  \BibitemOpen
  \bibfield  {author} {\bibinfo {author} {\bibfnamefont {V.}~\bibnamefont
  {Oganesyan}}, \bibinfo {author} {\bibfnamefont {S.~A.}\ \bibnamefont
  {Kivelson}}, \ and\ \bibinfo {author} {\bibfnamefont {E.}~\bibnamefont
  {Fradkin}},\ }\href {\doibase 10.1103/PhysRevB.64.195109} {\bibfield
  {journal} {\bibinfo  {journal} {Phys. Rev. B}\ }\textbf {\bibinfo {volume}
  {64}},\ \bibinfo {pages} {195109} (\bibinfo {year} {2001})}\BibitemShut
  {NoStop}%
\bibitem [{\citenamefont {Wu}\ \emph {et~al.}(2007)\citenamefont {Wu},
  \citenamefont {Sun}, \citenamefont {Fradkin},\ and\ \citenamefont
  {Zhang}}]{wu07}%
  \BibitemOpen
  \bibfield  {author} {\bibinfo {author} {\bibfnamefont {C.}~\bibnamefont
  {Wu}}, \bibinfo {author} {\bibfnamefont {K.}~\bibnamefont {Sun}}, \bibinfo
  {author} {\bibfnamefont {E.}~\bibnamefont {Fradkin}}, \ and\ \bibinfo
  {author} {\bibfnamefont {S.-C.}\ \bibnamefont {Zhang}},\ }\href {\doibase
  10.1103/PhysRevB.75.115103} {\bibfield  {journal} {\bibinfo  {journal} {Phys.
  Rev. B}\ }\textbf {\bibinfo {volume} {75}},\ \bibinfo {pages} {115103}
  (\bibinfo {year} {2007})}\BibitemShut {NoStop}%
\end{thebibliography}%

\newpage
\begin{appendix}
\begin{widetext}
\section{Supplementary Information}
In order to evaluate the flow equations for the couplings and the source terms, we shall make use of the following integral:
\be
\label{eq:0216d}
\int \frac{d\omega}{2\pi} \int_{\Lambda (1-d\ell)}^\Lambda 
	\frac{d^2 k}{(2\pi)^2} \hat{G}_0 (i\omega,\bk) \otimes
	\hat{G}_0 (\pm i\omega, \pm\bk) = \frac{m}{4\pi} d\ell \left[\mp A^{ph,pp} 1_4 \otimes 1_4
	+ \frac{1}{2} B^{ph,pp} \sigma_3 1 \otimes \sigma_3 1
	+ \frac{1}{2} C^{ph,pp}\sigma_1 1 \otimes \sigma_1 1 \right],
\ee
where $\hat{G}_0 (i\omega, \bk)$ is the bare Green function following from \eqref{eq:0108a}. The upper and lower signs on the left hand side of \eqref{eq:0216d} correspond to the labels $ph$ and $pp$ on the right hand side, respectively. The coefficients on the right hand side of \eqref{eq:0216d} are given by
\ba
\label{eq:0216c}
A^{ph} &= -\frac{\sqrt{2}}{\pi} \frac{K(1-\cot^2\eta)}{|\sin \eta |} \\
B^{ph} &= \frac{2\sqrt{2}[E(1-\cot^2\eta)-\cot^2\eta K(1-\cot^2\eta)]}{\pi|\sin\eta|(1-\cot^2\eta)} \\
C^{ph} &= \frac{2\sqrt{2}[K(1-\cot^2\eta)- E(1-\cot^2\eta)]}{\pi|\sin\eta|(1-\cot^2\eta)} .
\ea
and
\ba
\label{eq:0216e}
A^{pp} &= \frac{\sqrt{2}}{\pi |\sin\eta|} \left[ K (1-\cot^2 \eta ) 
	+ \frac{\lambda^2}{\sin^2\eta - \lambda^2} 
	\Pi \left( \frac{\sin^2\eta-\cos^2\eta}{\sin^2\eta - \lambda^2}, 1-\cot^2\eta \right) \right], \\
B^{pp} &= \frac{2^{3/2}}{\pi |\sin\eta| (\tan^2\eta - 1)} \left[ -K (1-\cot^2 \eta ) 
	+ \Pi \left( \frac{\sin^2\eta-\cos^2\eta}{\sin^2\eta - \lambda^2}, 1-\cot^2\eta \right) \right], \\
C^{pp} &= \frac{2^{3/2}}{\pi |\sin\eta| (1 - \cot^2\eta)} \left[ K (1-\cot^2 \eta ) 
	- \frac{\cos^2 \eta - \lambda^2}{\sin^2\eta - \lambda^2}
	 \Pi \left( \frac{\sin^2\eta-\cos^2\eta}{\sin^2\eta - \lambda^2}, 1-\cot^2\eta \right) \right].
\ea
Note that the coefficients in \eqref{eq:0216c}, which correspond to particle-hole scattering processes, are independent of the particle-hole asymmetry $\lambda$. The functions $K$, $E$, and $\Pi$ are complete elliptic functions. In the limit of particle-hole symmetry ($\lambda=0$) and rotational invariance ($\eta = \pi/4$), \eqref{eq:0216c} and \eqref{eq:0216e} reduce to $A^{ph,pp} = B^{ph,pp} = C^{ph,pp} = 1$.

From evaluating the one-loop diagrams shown in Figure \ref{fig:diagrams}, one obtains the coefficients $A_{ijk} = \sum_{m=1}^5 A_{ijk}(m)$ in the flow equation for the couplings [\eqref{eq:0128a} in the main text], where $m$ labels the five diagrams in Figure \ref{fig:diagrams}. 
One gets only diagonal contributions from the first diagram:
\be
\label{eq:0128c}
A_{iii}(1) =  \left[ -4 A^{ph}
	+ C^{ph}~\mathrm{Tr} ((\sigma_1 \sigma_j)^2)
	+ B^{ph}~\mathrm{Tr} ((\sigma_3 \sigma_j)^2) \right] \frac{m}{4\pi}.
\ee
From the second and third diagrams combined:
\be
\label{eq:0128d}
A_{iij}(2) + A_{iij}(3) = \bigg[ 
	A^{ph}~\mathrm{Tr}(\sigma_i \sigma_j\sigma_i \sigma_j)
	- \frac{1}{2}C^{ph}~\mathrm{Tr} (\sigma_i \sigma_j\sigma_1 \sigma_i \sigma_1 \sigma_j) 
	- \frac{1}{2}B^{ph}~\mathrm{Tr} 
	(\sigma_i \sigma_j\sigma_3 \sigma_i \sigma_3 \sigma_j) \bigg] \frac{m}{4\pi} .
\ee
From the fourth diagram:
\ba
\label{eq:0128e}
A_{ijk}(4) = \frac{1}{8} \bigg[ & 2A^{ph}
	~\mathrm{Tr}(\sigma_k \sigma_j \sigma_i)~\mathrm{Tr}(\sigma_j \sigma_k \sigma_i)
	- C^{ph}~\mathrm{Tr}(\sigma_k \sigma_1 \sigma_j \sigma_i)
	~\mathrm{Tr}(\sigma_j \sigma_1 \sigma_k \sigma_i) \\
	&\quad\quad- B^{ph}~\mathrm{Tr}(\sigma_k \sigma_3 \sigma_j \sigma_i)
	~\mathrm{Tr}(\sigma_j \sigma_3 \sigma_k \sigma_i) \bigg]  \frac{m}{4\pi} .
\ea
Finally, from the fifth diagram:
\be
\label{eq:0128e}
A_{ijk}(5) = -\frac{1}{8} \bigg[  2A^{pp}
	\left(\mathrm{Tr}(\sigma_k \sigma_j \sigma_i)\right)^2
	+ C^{pp} \left(\mathrm{Tr}(\sigma_k \sigma_1 \sigma_j \sigma_i)\right)^2 
	+ B^{pp} \left(\mathrm{Tr}(\sigma_k \sigma_3 \sigma_j \sigma_i)
	\right)^2 \bigg]  \frac{m}{4\pi} .
\ee
\begin{figure}
\centering
\includegraphics[width=0.7\textwidth]{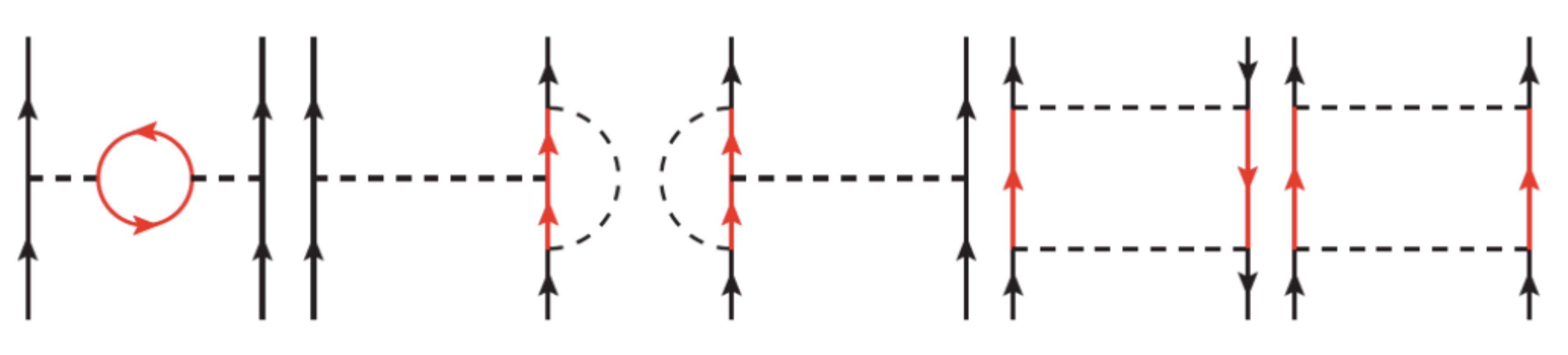}
\caption{Feynman diagrams contributing to the flow of the couplings $g_i$. The solid lines correspond to fermions; the dashed lines correspond to interactions. The internal fermion lines (colored red) have ``fast'' momenta $\Lambda (1-d\ell) < |\bk| < \Lambda$, while the external legs have ``slow'' momenta $|\bk| < \Lambda (1-d\ell)$.
\label{fig:diagrams}}
\end{figure}

The flows of the source terms introduced in \eqref{eq:0206a} are computed by evaluating the diagrams shown in Figure \ref{fig:diagrams2}.
\begin{figure}
\centering
\includegraphics[width=0.6\textwidth]{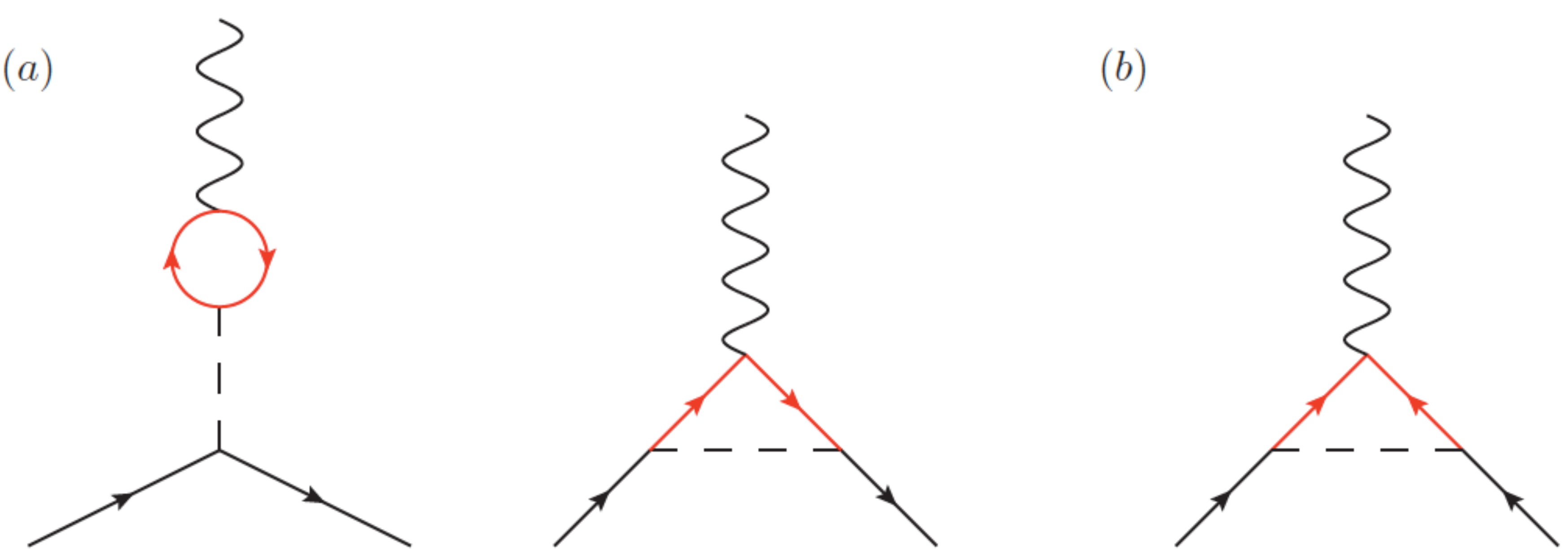}
\caption{Feynman diagrams contributing to the flow of the source terms $\Delta_i^{c,s,pp}$. The solid lines correspond to fermions; the dashed lines correspond to interactions. The two diagrams in (a) give contributions to the particle-hole source terms, while the diagram in (b) contributes to the particle-particle source term. The internal fermion lines (colored red) have ``fast'' momenta $\Lambda (1-d\ell) < |\bk| < \Lambda$, while the external legs have ``slow'' momenta $|\bk| < \Lambda (1-d\ell)$.
\label{fig:diagrams2}}
\end{figure}
Evaluating these diagrams and using \eqref{eq:0216d} gives the following expressions for the coefficients appearing in \eqref{eq:0206h}:
\ba
\label{eq:0220c}
B_{ij}^{(c,s)}(1) &= \frac{m}{4\pi} \{ - A^{ph} \mathrm{Tr} [(\sigma_j 1) M_i^{(c,s)}] 
	+ \frac{1}{2} B^{ph} \mathrm{Tr} [(\sigma_j 1) (\sigma_3 1) M_i^{(c,s)} (\sigma_3 1)]
	+ \frac{1}{2} C^{ph} \mathrm{Tr} [(\sigma_j 1) (\sigma_1 1) M_i^{(c,s)} (\sigma_1 1)] \}, \\
B_{ij}^{(c,s)}(2) &= -\frac{m}{16\pi} \{ -A^{ph} \mathrm{Tr} [((\sigma_j 1) M_i^{(c,s)})^2] 
	+ \frac{1}{2}B^{ph} \mathrm{Tr} [M_i^{(c,s)} (\sigma_j 1) (\sigma_3 1) M_i^{(c,s)} (\sigma_3 1) (\sigma_j 1)] \\
	&\quad\quad\quad\quad\quad\quad\quad\quad\quad\quad
	+ \frac{1}{2}C^{ph} \mathrm{Tr} [M_i^{(c,s)} (\sigma_j 1) (\sigma_1 1) M_i^{(c,s)} (\sigma_1 1) (\sigma_j 1)] \}, \\
B_{ij}^{(pp)} &= -\frac{m}{16\pi} \{ A^{pp} \mathrm{Tr} [(\sigma_j 1) M_i^{(pp)} (\sigma_j 1)^T  M_i^{(pp)}] 
	+ \frac{1}{2}B^{pp} \mathrm{Tr} 
	[ M_i^{(pp)} (\sigma_j 1) (\sigma_3 1)  M_i^{(pp)} (\sigma_3 1) (\sigma_j 1)^T] \\
	&\quad\quad\quad\quad\quad\quad\quad\quad\quad\quad
	+ \frac{1}{2}C^{pp} \mathrm{Tr} 
	[ M_i^{(pp)} (\sigma_j 1) (\sigma_1 1)  M_i^{(pp)} (\sigma_1 1) (\sigma_j 1)^T] \} .
\ea
Adding the contributions from the first two diagrams together then gives $B^{(c,s)}_{ij} = B^{(c,s)}_{ij}(1) + B^{(c,s)}_{ij}(2)$. By differentiating the free energy with respect to the source terms, one obtains the following expression for the particle-hole susceptibilities in the charge and spin channels:
\be
\label{eq:0220a}
\chi_i^{(c,s)}(\ell) = \frac{m}{4\pi} \int_0^\ell d\ell' e^{2\Omega_i^{(c,s)}(\ell')}
	\left\{ - 4 A^{ph} + \frac{1}{2} B^{ph} \mathrm{Tr} [(\sigma_3 1) M_i )^2]
	+ \frac{1}{2} C^{ph} \mathrm{Tr} [(\sigma_1 1) M_i )^2] \right\},
\ee 
where
\be
\label{eq:0220b}
\Omega_i^{(c,s)} (\ell) = \int_0^\ell d\ell' \sum_{j=0}^3 B_{ij}^{(c,s)} g_j(\ell').
\ee
The particle-particle susceptibilities are given by
\be
\label{eq:0220a}
\chi_i^{pp}(\ell) = \frac{m}{4\pi} \int_0^\ell d\ell' e^{2\Omega_i^{pp}(\ell')}
	\left\{ 4 A^{pp} + \frac{1}{2} B^{pp} \mathrm{Tr} [(\sigma_3 1) M_i )^2]
	+ \frac{1}{2} C^{pp} \mathrm{Tr} [(\sigma_1 1) M_i )^2] \right\},
\ee 
where
\be
\label{eq:0220b}
\Omega_i^{pp} (\ell) = \int_0^\ell d\ell' \sum_{j=0}^3 B_{ij}^{pp} g_j(\ell').
\ee

\end{widetext}
\end{appendix}
\end {document}